\documentclass[pre,aps,twocolumn]{revtex4}
\usepackage{graphicx}
\newcommand{\be}{\begin{equation}}
\newcommand{\ee}{\end{equation}}
\newcommand{\ba}{\begin{eqnarray}}
\newcommand{\ea}{\end{eqnarray}}
\begin{document}

\title{Kinesin as an electrostatic machine}
\author{A. Ciudad\dag, J.M. Sancho\dag and G.P. Tsironis\dag\ddag}
\title{Kinesin as an electrostatic machine}
\address{
\dag\ Departament d'Estructura i Constituents de la Mat\`eria, 
Facultat de F\'{\i}sica,
Universitat de Barcelona, Diagonal 647, E-08028 Barcelona, Spain\\
\ddag\ Department of Physics, University of Crete
and Institute of Electronic Structure and Laser, FORTH,
P.O. Box 2208, Heraklion 71003, Crete, Greece.
}
\date{\today}
\begin{abstract}
Kinesin and related motor
proteins utilize ATP fuel to propel themselves along
the external surface of microtubules in a processive and directional fashion.  We show that
the observed step-like motion is possible
through time varying charge distributions
furnished by the ATP hydrolysis circle while the static
charge configuration on the microtuble provides the guide for motion.
Thus, while the chemical hydrolysis energy induces appropriate local conformational
changes, the motor translational energy is fundamentally electrostatic.
Numerical simulations of the mechanical equations of motion
show that processivity and directionality are direct consequences of the ATP-dependent electrostatic interaction between the different charge distributions of kinesin and microtubule.
\end{abstract}

\maketitle

\section{Introduction}

The kinesin family is a set of motor proteins that move on the
surface of microtubules and shuttle various
cargo molecules to different parts of the cell\cite{Tuz}-\cite{parallel}.
The energy for kinesin
transport motion is provided by hydrolysis of ATP molecules that
are in its vicinity and attach to a specific site of the protein.
Despite much progress in the area, the precise, microscopic fashion of ATP
action is not known. Kinesin motion is step-like with one step for each ATP
hydrolysis while processivity and directionality depend on specific neck properties of each particular type of the family. These features of motor protein motion led several years ago in the introduction of mechanical ratchet models that could provide some
insight on the phenomenology of individual \cite{Reimann,Strat,Fisher,bier,sancho} and collective \cite{julicher,nishinari} motion.
Although many issues and especially those
related to the random aspects of the walk where addressed successfully by these
models, fundamental questions on the energetics and  nature of forces that enable
the walk have not been discussed.  In the present work we focus on the latter;
specifically we use electrostatic information of the microtubule and kinesin and
evaluate the full electric force that binds the latter on the former.  We then show that
the actual kinesin walk is caused by charge distribution changes on kinesin
enabled by the action of ATP.  The walk that is a combination of
conformational changes accompanied by kinesin charge reshuffling is found to be
fundamentally governed by electrostatic forces.
\section{Kinesin electrostatic model}
\begin{figure}[ht]
		\begin{center}
	\includegraphics{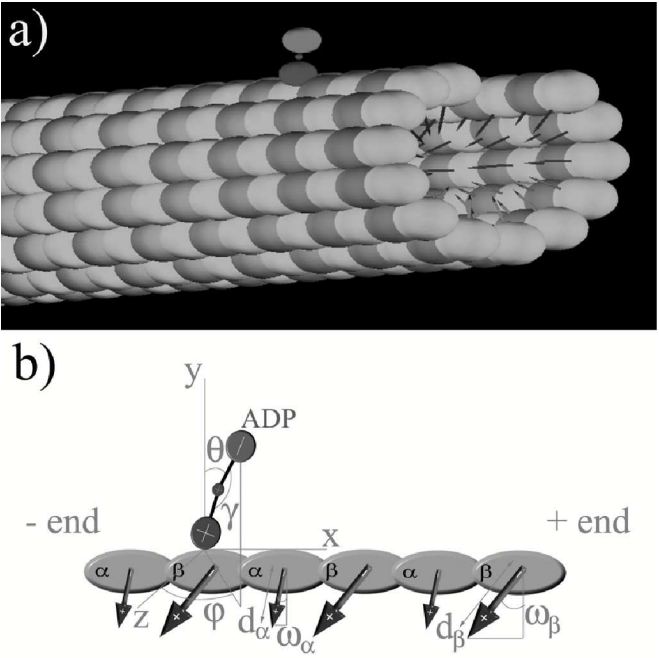}
	\end{center}
\caption{ a) Microtubule electrostatic model with arrows indicating local dipole moments.
The $\alpha$ tubulin subunits (dark) have smaller dipole moment that
the  $\beta$ units (lighter color).
b) Electrostatic configuration of kinesin and tubulin made protofilament prior to ATP
hydrolysis.  Angle $\theta$ is polar while $\phi$ is azimuthal.
The central-neck charge sign depends on the type of molecular protein while the head
charges depend on the ATP hydrolysis circle. Dipolar lenghts $d_{\alpha}$, $d_{\beta}$ and dipolar angles $\omega_{\alpha}$, $\omega_{\beta}$ are different
in $\alpha$ and $\beta$ subunits respectively. }
\label{1}
\end{figure}

Kinesins walk on microtubules; the latter are made by
alpha and beta of  tubulin  proteins that
are highly charged.  Molecular modeling shows that, in the presence of
water, tubulins have additionally a permanent dipole moment that points
towards the cylindrical symmetry axis of the microtubule with
a non-zero component parallel to the protofilaments towards their minus
end\cite{Tuz}.  Electrostatically thus a microtubule has negative surface
charge accompanied by a positive charge distribution in its immediate interior
with a slanted polarization vector that is larger in the beta than the
alpha subunit (Figure \ref{1}).  This symmetry breaking induced by the tubulin dipole moment direction
plays a key role in the specificity of the attachment of a kinesin head onto a
beta tubulin subunit\cite{decoration}.\\

Structural analysis of kinesin and related proteins suggests
that there are three charge domains,
one in each head region and a third
one in the central neck linker domain. The head charge distribution  depends strongly
on the ATP or ADP presence while the neck charge is specific to the protein.
For modeling purposes it is sufficient to consider kinesin and similar motor proteins
as a body containing three charges located on the apexes of a triangle with angle
$\gamma$ subtending from the neck charge to the two head charges (Figure \ref{1}b)
These charge distributions interact with tubulin multipole fields and determine
at each instant of time the dynamic state of the molecule.


In our analysis of kinesin walk to be detailed below, we place
motor proteins and microtubules in  an overdamped medium with thermal energy
$k_BT=4.1pNnm$ and motor drag force  $10^{-6}pNs/rad$ per rotational degree of freedom. For the sake of simplicity, we assume a relative electric permitivity equal to 80 everywhere and a Debye length $l_D$ above $3nm$. Although Debye-H\"uckel theory predicts $l_D\sim1$, the interaction range may be enhanced by the channeling of the electric field along the interior of the proteins, which is not accesible by diffusing ions or water molecules. Then it is reasonable to take $l_D$ of the order of the size of the kinesin. 
We take consecutive tubulin unit distance to be $8nm$, $4nm$ per $\alpha$ or $\beta$ subunit. The lateral space between two protofilaments is $5nm$  while the  pitch of $1nm$
between consecutive protofilamens is also included.  For a processive plus-ended kinesin
and its non-processive-minus-ended chimera ncd we use for the head-to-head distance the value of $6 nm$ while we consider the heads to be charged with $+2 e$ in the absence of nucleotides, $-2e$ with ATP attachment and $-1e$ with ADP.
These head charge values are consistent with the charges involved in hydrolysis reaction
\be
ATP^{4-} + OH^{1-}\rightarrow ADP^{3-}+H_2PO_3^{2-}.
\ee

The value of the central motor charge changes for different proteins; as a result
we consider different motors with charges in the range $[-2e:+2e]$.

For the electrostatic distribution of microtubules we assign negative surface
charge $q= -27 e$ per tubulin subunit while include a positive charge distribution
in the interior leading to dipole moment magnitude of
$ p = 5000D \simeq 100 Cnm$\cite{Tuz},
or $d \simeq 4nm$ ($p=qd$). Finally, we use a
dipolar tilt with length and angle values equal to
$d_{\alpha}\simeq 2nm$,  $\omega_{\alpha}\simeq 0.07 rad$
and $d_{\beta}\simeq 4nm$,
$\omega_{\beta}\simeq 0.14 rad$, for the $\alpha$ and $\beta$
subunits respectively (Figure \ref{1}b).

The microtubule-induced kinesin interaction potential is
given by
\be
V(\vec{r}_i)=\frac{-1}{4\pi\epsilon_0\epsilon_r(1+ka)}\sum_{j=1}^{N}
\frac{q_iq_j}{|\vec{r}_i -\vec{r}_j|}e^{k(a-|\vec{r}_i -\vec{r}_j|)},
\label{eq2}
\ee
where $\vec{r}_i$ is the position vector labeling the charges on kinesin
($i=1,2,3$), while $\vec{r}_j$ is the location of the N microtubule charges $q_j$
on the $\alpha$ and $\beta$ subunits. $k$ is the inverse of the Debye length, which we take around $\sim3.5 nm$, and $a\sim1 nm$ is the excluding volume radius as described in Ref. \cite{Levin}. A regime with a greater $a$ and lower Debye length $l_D$ is also operative.  
We considered a flat microtubule with five protofilaments; due to the rapid decay of the force out from protein volumes, we include in total the $N=10$ closest tubulin charges to kinesin. For the simplest case when neck and head charges are aligned we have
$\gamma=\pi$  and the protein reduces to a triply charged rigid rod.

\begin{figure}[ht]
			\begin{center}
	\includegraphics[width=8 cm]{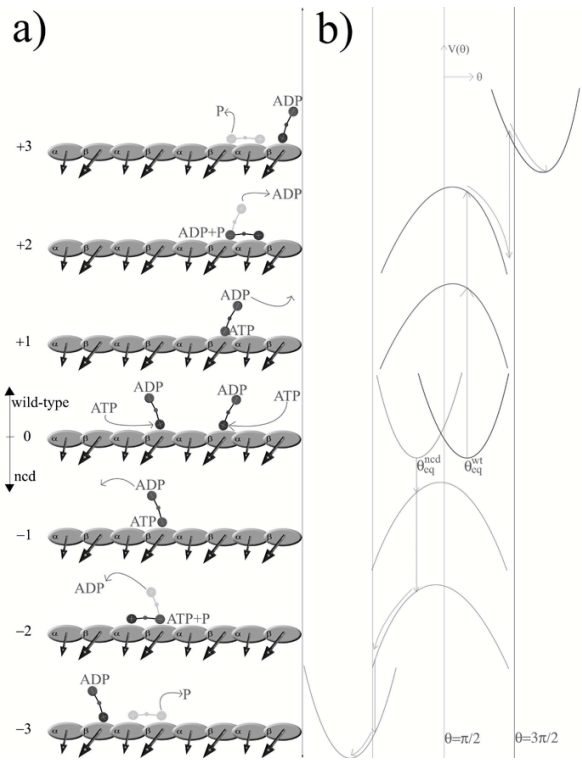}
	\end{center}
\caption{ Electrostatics driven motor walk:
Wild-type kinesin (+ states) and ncd (- states) stepping process (left column)
and numerically determined
binding protein-microtubule electrostatic potential as a function of local polar angle
$\theta$ (right column).
(0) Before ATP hydrolysis, both kinesin (positive neck) and ncd motors (negative neck) are in parked configuration pointing in opposite directions due to the difference in the central charge.  The corresponding
equilibrium angles are determined for the minimum potentials.
(+1) ATP entry in the kinesin (attached) head pocket, with an accompanying charge
change while the ADP at the other (tethered) head becomes unstable.
(+2) The  reversal and shift of the
interaction potential of the previous state leads to
falling of the tethered head deterministically towards the plus end. Since
the length of the motor is not sufficient for reaching the next tubulin subunit we
have (+3) a detachment and rising of the trailing head in such a way that allows the other head to slide to the next binding site.
The ncd motor protein cycle proceeds similarly (- states) but the parked state is tilted towards the minus-end. Moreover, the negative-central charge induces a potential shift which is opposite to the plus-ended case. The falling of ncd motors is slower than positive-charged-neck motors, so the probability that the attached heads begins the rising before the tethered heads completes the falling is greater, leading to non-processivity.}
\label{2}
\end{figure}

We have
tested the model for several values of $\gamma$ smaller than $\pi$; we
have found that the model is not fully opperational for
$\gamma \lesssim  175^0$.  We have also tested the model 
considering that the angle $\gamma$ has some elasticity as well.
We found that in case the angle is very soft,
the raising head rapidly loses its power to pull the motor forward. 
For the purposes of the model, the motor core and the neck need to be stiff enough
to maintain the values of $\gamma$ in the range discussed above even 
at maximum load conditions.
In this simpler rod-like configuration  the polar angle $\theta$ and the azimuthal angle $\phi$ are
sufficient for describing the motor rotation.   We can describe the motion using 
 the following overdamped equations of
motion for kinesin:

\be
\lambda\dot{\theta}=-\frac{1}{L}\frac{dV(\vec{r})}{d\theta}+\xi_{\theta}(t)
\label{eq3}
\ee
 and
\be
\lambda\dot{\phi}=-\frac{1}{L}\frac{dV(\vec{r})}{d\phi}+\xi_{\phi}(t)
\label{eq4}
\ee
 where $\lambda$ is the drag coefficient, $L$ the head-to-head distance
and
$V(\vec{r})$ the total microtubule electrostatic potential of Eq. (\ref{eq2}) at the
Cartesian location $\vec{r}$.
The environment is simulated through the thermal forces
$\xi_{\theta}(t)$ and $\xi_{\phi}$;  for each we have
$\langle\xi(t)\rangle=0$ and $\langle\xi(t)\xi(t')\rangle=2\lambda
k_BT\delta(t-t')$.   In order to integrate Eqns. (\ref{eq3}, \ref{eq4})
we need to perform at each instant of time the Cartesian-to-polar
transformation
$\theta=\theta(\vec{r},L)$.   For the rotation we consider the
attached head to coincide with the origin of the coordinate system;
the latter is shifted by $8nm$ each time a step is completed.  The simple
Larmor-like rotation of the protein for $\gamma = \pi$ becomes a
more complex rigid body rotation for $\gamma<\pi$.
Extensive numerical simulations with the model described previously for
the kinesin-microtubule complex led to the following quantitative picture for the motion.
In equilibrium (parked) state the nucleotide-free head  is  positively
charged and thus attached  to the beta subunit of the negative microtubule surface
while the other head containing ADP is negatively charged and tethered through the neck.
Due to the slanted microtubule dipole moment, the kinesin axis is tilted but in a direction
determined by the central charge sign; for wild type kinesin
points to the fast-growing-end while for ncd points to the minus end ( Figure \ref{2}).

All experimental data agree that
after nucleotide entry in the head pocket, the binding of the attached head to the microtubule
and the ADP binding to the tethered head become unstable, signaling the onset of a cascade of events leading to the walk.
In our model this critical juncture occurs because in the absence of nucleotide the positively charged head becomes negative
acquiring $-4e$ charges after ATP  binding.  Although the entry of a negatively
charged molecule in the head pocket is necessary for the commencement of the walk
process, the release of the attached head is not immediate, since a head-tubulin chemical bond needs to be broken. ATP hydrolysis energy thus is used for
a local conformational change that captures the negative charge after
dephosphorylation.  Although charged ADP nucleotides in the medium compete with ATP entering also in the pocket, they do not hydrolyze and, as a result, cannot induce the local conformation that
will trap them. In our electrostatic model thus hydrolysis energy induces
primarily a specific local conformational change, being the power stroke electrostatic in nature.
\section{Kinesin dynamics}

Based on this picture, the protein parked configuration
is a result of the microtubule-empty head attraction, microtubule-tethered
head repulsion while the kinesin angle wrt microtubule
is due to the tubulin dipolar tilts.  This stable configuration is
maintained during a random and [ATP]-load-dependent dwell time
until a new ATP binds into the attached head.
When this occurs, the local charge changes
and the ADP trapped in the tethered head experiences an additional repulsion,
becoming unstable and eventually opening the pocket and exiting.
After ADP expulsion the tethered head becomes positive, the attached head charge is negative
and the protein becomes electrostatically unstable.  As a result the tethered head
collapses onto the microtubule ("falling processes"), while the attached head
is forced to detach from the surface ("rising process") and a new, shifted, park state is
reached.

Every kinesin step thus includes
falling of the leading head onto the microtubule followed by
rising of the trailing head ( Figure \ref{2}).
We performed extensive  3D numerical simulations and found that the
electrostatic force field generated by the microtubule is able to
drive these two sequential  processes while keeping the motor faithful to
a given protofilament.
We note that, energetically, the parked state represents a potential energy minimum for the
interaction between free head charge and microtubule.
When ADP is released and the electrostatic charge distribution of the
motor changes this   minimum becomes a potential
maximum, i.e. the repulsion of the tethered head turns into attraction.

For a null-charged-neck case, even though the motor is tilted in the parked state, there is no motion directionality since the falling of the tethered head does not have
a preferred collapsing side. The directionality features of the motor enter through the
charge distributions of the neck region, which is responsible of the potential shift between stable and unstable equilibrium points. Specifically, when the central charge
is positive, not only the protein is more processive due to the attraction
with the negatively charged tubulin surfaces, but it also walks towards
the plus-end of the microtubule; this is precisely the case with wild-type kinesin.
In ncd, on the other hand, the central charge is negative, the protein is
non-processive and walks in the opposite direction towards the minus end.
Additionally, in proteins with no charge in the neck region,
the walk is known not to be deterministic but random\cite{Mutante}.
We note further that our simulations
are compatible with the experiments in reference \cite{parallel}
whereby the
falling motor direction is parallel to the protofilament axis, with
some variations and latteral collapses leading to protofilament changes
due to thermal fluctuations.
The lateral periodicity of the microtubule lattice is $5nm$ while the axial
one only $4nm$; these values determine crucially
the motion after the destabilization of the parked state since the electric force is stronger in the axial direction.
We find that this electrostatic model captures fully the directionality and
processivity features of the known motors.

\begin{figure}[ht]
	\begin{center}
	\includegraphics{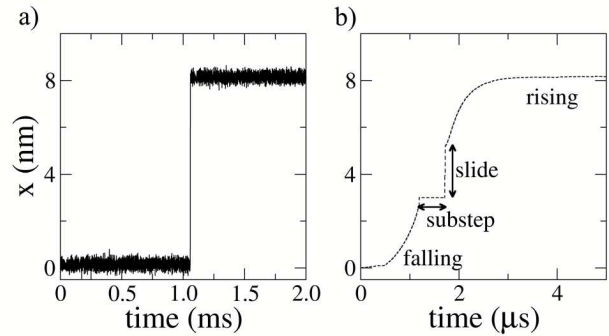}
	\end{center}
\caption{Numerical results for a kinesin step at (a) $ms$ time-scale and (b) $\mu s$
time-scale, for  $\gamma=\pi$; $x$-axis denotes time while $y$-axis kinesin
position..
Occurrence of non-processivity and substeps are in competition since for the former
detachment of the trailing head occurs before than the falling of tethered head, while for substeps to occur the falling process must be done before detachment, leading to a processive motion.
The millisecond time-scale numerical results shown in (a) are detailed
at a microsecond time-scale in (b) where the
aforementioned competition is seen.
}
\label{3}
\end{figure}

Two different time-scales characterize the actual kinesin motion.
While dwell times between
two steps, modulated by ATP concentration  and the external load
are of the order of milliseconds\cite{Visscher},
the action of the step itself lasts only some
microseconds \cite{substepsno}; this is the reason for the stepping
appearance of the trajectories.
In the relatively short period of the step time, several
processes must occur. First, falling of the leading head
after ADP release, a process activated by ATP binding on the attached head.
Subsequently the two heads remain attached onto the  microtubule for a 
time not longer than 30 microseconds \cite{substepsno}.
Recent measurements were able to track bead-movements at the
microsecond scale \cite{substepsno}. The recording of a single step, in our interpretation, reveals two different regimes that may be separated by a quick stage at which both kinesin heads are attached to the microtubule.
Kinesin crytallization data
show that the separation
between heads is smaller than 8 nm\cite{cristalina}, i.e.
slightly smaller than
the protofilament spatial period and, as a result, 
after falling the leading head
cannot reach the beta subunit location before the trailing head detaches.
Thus, sliding may take place while  the trailing head rises to reach
the new parked state.  This aspect is portrayed in
Figure \ref{3} in a sequential form although in practice both
sliding and rising could occur simultaneously.

In order to address motor protein  processivity
we consider the two sets of chemical reactions that are activated by ATP binding at the trailing head. The first group is related to a cascade of rapid reactions  involving
ATP hydrolysis, dephosphorylation and subsequent head detachment
with an estimated time 500
$\mu s$ for each\cite{chemistry}.
Secondly, ATP  activates ADP release in the other head, although
the typical time of this process is not known. There is a
competition between these two reactions; if hydrolysis and
detachment occur faster, the whole motor detaches from the microtubule
and processivity ceases. If, on the other hand, ADP release and falling is the fastest
process, then a small substep occurs while both heads are attached.
Processivity thus is related to the competition of these two processes.
In the case of non-processive  ncd, for instance, the existence of a
negative central  charge has the effect of slowing
down the falling process while
speeding up the rising one resulting, thus, in a non-processive motor. This effect has been tested experimentally in Ref. \cite{Vale}.\\

We point out that when ATP enters into the attached head, 
the system becomes electrostatically unstable. 
However, detachment of the head occurs some time later triggered by
the phosphate release. This fact implies that the chemical bonding 
that binds the motor to the microtubule,  while
stable in absence of ATP, becomes unstable when the 
nucleotide arrives. The time delay between ATP entry and bond dissociation
appears to be crucial for motor processivity. 
In the Figure \ref{2}, we have assumed that both possesive and 
non-possesive motors may  perform the full detachment-attachment cycle, 
although, in the case of ncd, the chemical bond may be actually  broken 
before the falling head reaches the next binding site.
The time duration of a single kinesin step
is very small compared to the parked dwell times, and, as a result,
global observables such as  the mean velocity or randomness 
can be predicted using only chemical kinetics. 
This has been shown in  reference \cite{Ciudad} where a 
kinetic approximation was shown to be sufficient
in fitting the measured data, even in the presence of an external load. 
However, our present approach shows, that thermal fluctuations alone
may not produce directly the $16 nm$ long displacements in a $\mu s$ time-scale
necessary for the walk.

\section{Conclusions}

In this work we attempted to analyze the complex motion of molecular motor 
proteins from  a mesoscopic point of view placing emphasis on the fundamental
interactions that enable the motion.  We found that the motion is primarily driven
by the electrostatic interaction between the charged microtubule surface and
the fluctuating motor head charges.  The nonequilibrium aspect of the walk is 
provided by the ATP hydrolysis cycle that furnishes appropriate
charge motor distributions that make the walk possible.  We made several 
assumptions in this work, most of which have been motivated by the
current status of knowledge in the area.  We considered the motor as a relatively
rigid one; this assumption is presently well founded, 
however it may be lifted through an
improved motor model that includes additionally protein eleasticity.
Furtheremore, our model does not address at all the structual changes
in the head pockets effected by ATP hydrolysis, the mode of the
local energy relase as
well as the mechanism for the  ADP detachment from the thethered head.  
These issues that may 
involve  complex conformational changes as well as possibly charge or energy 
transfer processes must be addressed through a more foundamental, 
microscopic model.  Finally,
it is crucial to evaluate the range of Debye screening in the vicinity of 
a highly charged surface such as the microtubule.  While our motor model
remains fully operational to a Debye length range of approximately $3 nm$, one 
certainly needs to address the complex nature of the electrostatic shielding in the
vicinity of the microtubule and  assess the true range of the electrostatic
forces involved in the walk.
We note that our model provides a consistent dynamical 
picture for the kinesin walk  that is
based on two premises, viz. 
(a) that the ATP hydrolysis energy is used for a head
local conformational change that captures locally charges and  (b) the motor
motion is driven by electrostatic forces.  The emerging qualitative and
quantitative picture for the walk is fully compatible with all known 
experimental data, while, furthermore, it is testable experimentally.
If electrostatics, charge transfer as well as capture
indeed link the chemistry of ATP and the mechanics of 
kinesin, this may hold true for other ATP-dependent processes as well.

\acknowledgements
We thank Marta Iba\~nes for discussions.
This work was supported by the Ministerio de Educaci\'on y Ciencia
(Spain) under Project No. $FIS2006-11452-C03-01$, Grant No. $BES-2004-3208$,
by grant $2006PIV10007$ of the Generalitat de Catalunya
and Grant "Pythagoras II" $KA-2102-TDY-25$ of the Ministry of 
Education of Greece
and the European Union.

\end{document}